\makeatletter                                                           
\@ifundefined{epsfbox}{\@input{epsf.sty}}{\relax}                       
\def\eps@scaling{.95}                                                   
\def\epsscale#1{\gdef\eps@scaling{#1}}                                  
\def\plotone#1{\centering \leavevmode                                   
\epsfxsize=\eps@scaling\columnwidth \epsfbox{#1}}                       
\makeatother                                                            
\def\plotbig#1{\centering \leavevmode
\epsfxsize=1.9\columnwidth \epsfbox{#1}}
\documentstyle{mn}

\title{Spiral Shocks in 3D Accretion Discs}
\author[]
       {Hiroshi Yukawa, Henri M.J. Boffin\thanks{Present address: Dept. of Physics and Astronomy, 
University of Wales, Cardiff CF2 3YB, Wales} and Takuya Matsuda\\
        Astrophysics Laboratory, Department of Earth and Planetary Sciences, Rokkoudai 1-1,
        Nada ku, Kobe 657, Japan}
\date{Accepted July 21, 1997. 
      Received June 1, 1997; 
      in original form February 10, 1997}

\begin{document}

\maketitle

\begin{abstract}
We have performed three-dimensional numerical simulations of accretion discs
in a close binary system using the Smoothed Particle Hydrodynamics (SPH) method.
Our results show that, contrary to previous claims, 
3D discs do exist even when 
the specific heat ratio of the gas is as large as  
$\gamma=1.2$.
Although the disc is clearly more spread in the $z$-direction in this case than it is
for the quasi-isothermal one, the disc height is compatible with the hydrostatic 
balance equation.

Our numerical simulations with $\gamma=1.2$ also demonstrate that spiral shocks exist in 
3D discs. 
These results therefore confirm previous 2D simulations.

\end{abstract}

\section{Introduction}
Numerical studies of accretion discs have been
mostly restricted to 2-dimensional cases, due to computing time limitations.
Among many things, these 2D simulations have shown that spiral shocks 
appear in inviscid discs 
(Sawada et al. 1987, Rozyczka \& Spruit 1989; Spruit 1989).
These spiral shocks have been invoked as an alternative to the $\alpha$-viscosity
theory for transfering angular momentum outward, thereby allowing accretion.
It is however still unclear whether these spiral shocks 
are effective enough to explain observations. 

Recently, some 3D simulations have been carried out.
Except for Sawada \& Matsuda (1992), they were
all performed using particle methods 
(Molteni, Belvedere \& Lanzafame 1991; Hirose, Osaki \& Mineshige 1991;
Lanzafame, Belvedere \& Molteni 1992, 1993;
Meglicki, Wickramasinghe \& Bicknell 1993;
Whitehurst 1994; Simpson 1995; Armitage \& Livio 1996). 

Sawada and Matsuda (1992) used the Eulerian grid based Roe upwind TVD scheme
to calculate the case $\gamma=1.2$ and mass ratio $M_1/M_2=1.0$ 
up to half an orbital period.
They found the accretion disc to be not axisymmetric but characterized
by a pair of spiral shocks, although their computational time was not 
long enough to reach a steady state.

Hirose et al. (1991) used the "sticky" particle method of Lin \& Pringle (1976). In this 
method, the effect of pressure is neglected but not viscous interactions, and the energy generated
is assumed to be radiated away instantaneously. Hirose et al. (1991) considered two values
of the mass ratio, 0.15 and 1. They observed that the vertical height of the disc, $H$, is generally
much larger than the value expected from the hydrostatic balance equation. The height of
the disc also appeared to be a variable function of the orbital phase and of the distance ($r$). 
They attributed this to the interaction between an incoming stream and the gas in the disc.
The ratio $H/r$ they obtained was typically between 10 and 15 \%.

Molteni et al. (1991) carried out 3-D simulations of the formation and 
the evolution of polytropic accretion discs using the SPH method. 
They made calculations for three values of the polytropic index, that is
 $\gamma=1.01$, 1.1 and 1.2, 
using 9899, 2372 and 1579 SPH particles respectively.
They concluded that the formation of a disc is inhibited for $\gamma\ge1.2$ and 
that the thin disc approximation is confirmed only for low $\gamma$ values.
Later, Lanzafame et al. (1992) performed calculations with a mass ratio $M_1/M_2=1.625$.
They concluded that disc formation is inhibited for $\gamma\ge1.1$, 
a value still lower than that found in the previous paper.

Meglicki et al. (1993), in an attempt to study large-scale eddies, 
also used SPH but with a very low value for the artificial viscosity.
Their model had been evolved only for 1.7 orbital periods 
but showed the presence of turbulence which may play a role in transporting
angular momentum. Turbulent eddies were indeed seen up to scales approaching half 
the thickness of the disc.

Whithehurst (1994) studied the eccentric disc model of superhump formation in SU UMa stars
using a free Lagrange method which approximates the gas as a collection of fixed-mass 
cells with variable volumes. He typically used about 60 000 cells in his 3D calculation
of accretion disc with a mass ratio smaller than 0.333 . An ad-hoc artificial viscosity
is included in his method to allow for angular momentum transfer in a way comparable to 
the standard $\alpha$-disc theory. Energy dissipated was allowed to radiate away instantaneously.

Simpson (1995) calculated the case $\gamma=1.01$ and $M_1/M_2=12.5$ 
until about 50 orbital periods with SPH, using 5 000 particles.
No eccentric modes were seen to develop.
However, spiral density waves were visible for most of the later orbits.

Armitage and Livio (1996) investigated the collision between the stream from the
inner Lagrange point, L$_1$, and
the accretion disc using the SPH method, in a binary system with a mass-ratio of 0.2. 
They used an isothermal equation of state and set-up initially the disc as an annulus of 30,000
particles. 
They observed that significant amounts of stream material overflowed or
bounced off the edge of the disc and flowed over the disc surface toward smaller radii.
This, they believe, is 
a likely cause of the absorption dips observed  
in some nearly edge-on low-mass X-ray binaries.

One of the most striking facts from most of these particle calculations is
their apparent inability to generate spiral shocks in the accretion discs. 
This may be related to the fact that, except for Molteni et al. (1991) and Lanzafame 
et al. (1992), the authors considered isothermal or pseudo-isothermal cases or
they used a large amount of viscosity in their method. 
Therefore, these calculations do not constrain very much
the formation of spiral shocks in 3D.
On the other hand if spiral shocks were to exist in 3D, the calculations of Molteni et al. (1991)
and Lanzafame et al. (1992) should have shown them. However, as already discussed in 
Boffin, Yukawa \& Matsuda (1997; hereafter Paper I), 
their use of a very small number of particles in the $\gamma = 1.1 $ and 1.2
cases, of an arithmetic mean for the pressure gradient 
as well as of a constant smoothing length, may lead to incorrect results.

We would like therefore, in this paper, to address two questions :
(i) is disc formation really inhibited for value of $\gamma \ge 1.1$ ? 
(ii) Can we extend the results of Paper I and also observe spiral shocks in 3D ?

\section{Method}
We use the SPH method (see e.g. Monaghan 1992 for a review),
which is rather well suited to the study of mass transfer in a binary system.
The details of our numerical method are described in Paper I.

We choose the following units: orbital separation
$ A $, total mass $ M = M_1 + M_2 $ and orbital
velocity $ V_{orb}=[G(M_1+M_2)/A]^{1/2} $.
The orbital period is therefore normalized to $2\pi$.
We will study only cases with mass ratio equal to unity, 
i.e. $M_1=M_2=0.5$.

We assume a non-viscous gas and use  
the standard SPH artificial viscosity with $\alpha=1.0$
and $\beta=2.0$. We do not take into account any radiative cooling effect.
We work in the non-inertial frame of reference where both stars are at rest. 
The primary is centered at the origin, the secondary at (-1,0,0).

We carry out two kinds of simulations : with and without
mass supply from the secondary star. In both cases, the computing region was a
sphere of 0.55 around the accreting star whose radius is 0.03. 

In the Roche lobe overflow(RLOF) case, gas from the secondary 
star flows with the velocity of sound through the L$_1$ point toward the primary. 
The sound speed of the gas is 0.1 and its density is 1. 
SPH particles at the L$_1$ point are ordered on a square lattice of width $0.04$.
Throughout the calculation, we keep a constant value of the injection rate, $ 1.6 \times 10^{-4} $.
The initial value of the smoothing length is 0.0026 ($\gamma=1.2$), 
0.00308 ($\gamma=1.1$) or 0.005 ($\gamma=1.01$). At the end of the simulation, 
it ranges from 0.0026 to 0.85, with a mean value of 0.0067 ($\gamma=$1.01) or 0.0094 ($\gamma=$1.2).

\begin{figure}
  \plotone{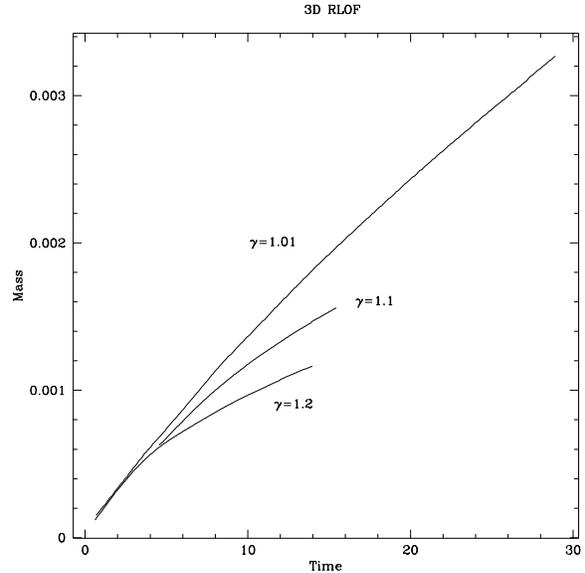}
  \caption{Evolution of the disc mass as a function of time, for the 3
simulations of Sect. 3.1}
  \label{masstime}
\end{figure}
 
In the second kind of simulation, we calculate the evolution of the disc from an initial state
in which each particle has been given a Keplerian velocity. Here, only the case $\gamma=1.2$ has
been investigated.
The radius of the initial disc is 0.30 and the thickness is 0.08. The gas in the 
disc has also a sound speed of 0.1 and a density of 1. 
The vertical structure of the disc is obtained by placing 9 layers - each separated by 0.01 - on
top of each other. This ensures that the disc is resolved vertically. 
Very quickly, the disc adjusts itself so that the particles
distribution in the $z$-direction is gaussian (see also below).
Each layer consists of an annulus between $r=0.05$ and $r=0.30$. Particles are placed
on concentric rings such that the mean distance between particles is 0.01.
Initially, the smoothing length is 
0.011 and there are 31 000 particles. At the end of the simulation, at about T=2.3, the 
mean value of the smoothing length for the remaining 18,000 particles is 0.026 .

  \begin{figure*}
    \caption{The time evolution of the particle location map in the orbital plane
in the case of $\gamma=1.2$.}
    \label{g12time}
  \end{figure*}
 
\setcounter{figure}{1}
  \begin{figure*}
    \caption{Continued}
    \label{g12time2}
  \end{figure*}

\section{Results}

\subsection{Roche lobe overflow}

We have calculated three cases corresponding to $\gamma=1.2$,$1.1$ and $1.01$,
until $T \simeq 14$ (that is, a little more than 2 orbital periods),
$T \simeq 19$ and $T \simeq 39$, respectively.
As can be seen from Fig. 1, which shows that the disc masses
are still increasing, we have not yet reached a steady state.
It is interesting to note that the pressureless calculations of Hirose et al. (1991) 
reached a steady state after 
only about 100 orbital periods, while Molteni et al. (1991) get a stationary state after three
orbital periods or less. 
Our 2D results (see Paper I) show also that the mass of the disc is stationary only after
several tens of orbital periods. However, from these 2D results, as well as from the fact 
that the gross features in the flow do not change very much, we are confident 
that we have almost obtained the final structure of the disc.
 
The time evolution of the flow is shown for the $\gamma=1.2$ case
in Fig.~\ref{g12time}. Only particles within 0.02 of the orbital plane are
represented.  It is clear that the structure of the flow does not change after, say,
one orbital period. 
This was also observed in our 2D results. But, although the basic 
pattern is unchanged, the density in the disc increases continuously
as a result of an increase of the number of particles.
From Fig.~\ref{g12time} it is also obvious that
spiral shocks, similar to those seen in 2D, are present in the orbital plane. 

\begin{figure}
  \plotone{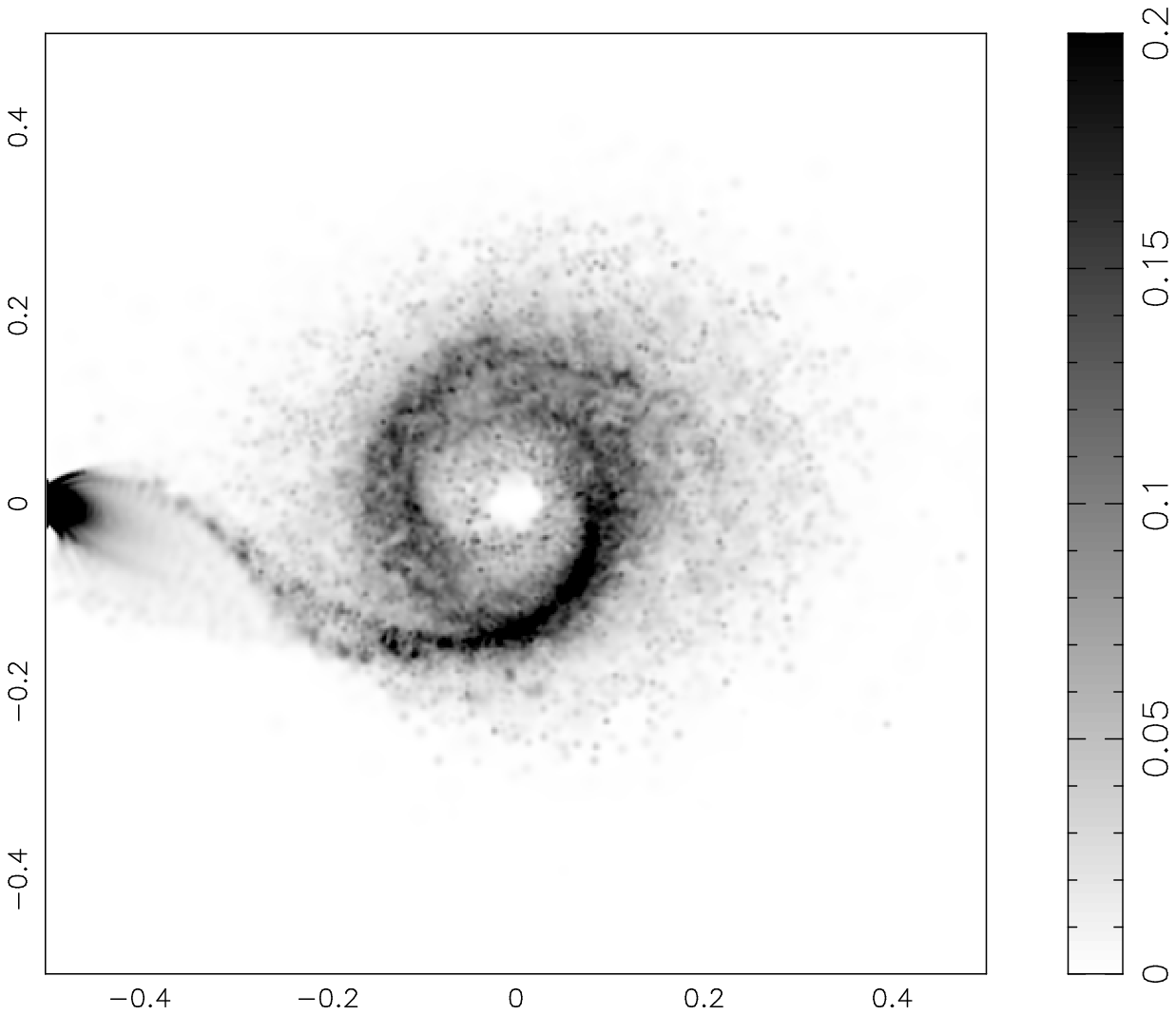}
  \caption{Gray scale plot of density in the $x-y$ plane at time $T\sim14$ 
in the case of $\gamma=1.2$.}
 \label{den.g12}
\end{figure}

We show our final results for the $\gamma=1.2$ case in Fig.~\ref{den.g12} and following.
At this stage, we have 61369 particles. The radius of the disc is approximately 0.2,
similar to the 2D results. 
Fig.~\ref{den.g12} clearly shows that the disc is asymmetric in the orbital plane: 
the upper half is more spread out than the lower one. 
The disc is however totally symmetric perpendicular to
the equatorial plane, as shown by Fig.~\ref{g12xz}.

\begin{figure}
  \caption{The location of particles in the $x-z$ plane at time $T\sim14$
in the case of $\gamma=1.2$ : 13730 particles out of a total of 61369 are shown.}
  \label{g12xz}
\end{figure}

\begin{figure*}
  \plotbig{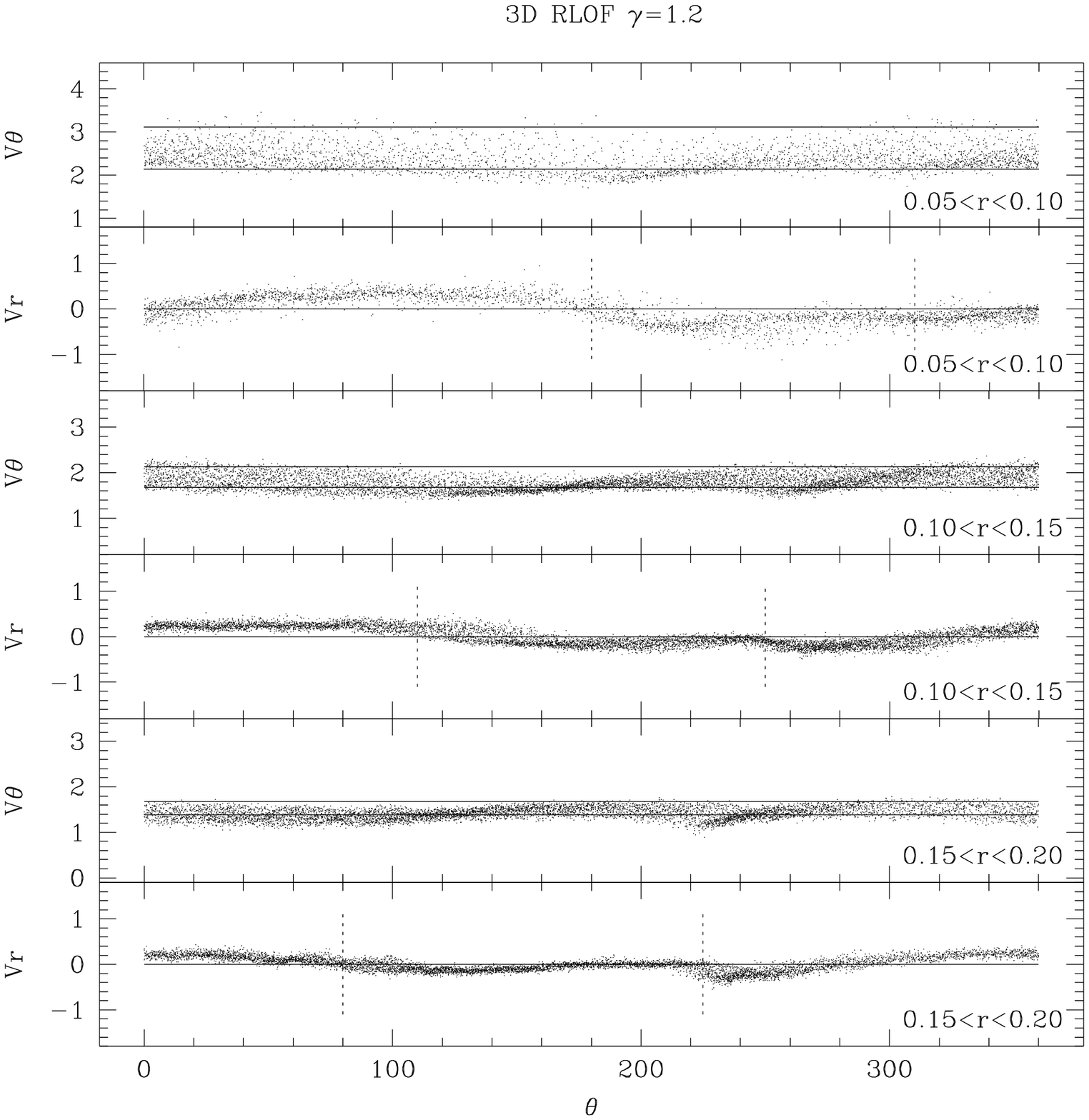}
  \caption{Azimuthal plots of the azimuthal and radial components of the velocity for the case $\gamma=1.2$
at time $T \sim 14$.
Only particles with $|z|<0.02$ are shown. The two solid lines in the azimuthal velocity plots represent the
Keplerian velocity corresponding to the inner and outer radius of each region.
The dashed lines indicate the approximate positions of the shock.}
  \label{av.g12}
\end{figure*}

In order to help understand the flow structure, we analyse now in more detail the
velocity profile. First, in Fig.~\ref{av.g12}, the azimuthal and radial components of
the velocity are shown as a function of the azimuthal angle, $\theta$, for three different
ranges in radial distance.
For the azimuthal velocity, the two horizontal solid lines represent the Keplerian velocity
corresponding to the inner and outer radius of each region. In the radial velocity plots, the 
dashed lines indicate the approximate positions of the shock. Figure ~\ref{av.g12} clearly
demonstrates the effect of the shock : the azimuthal component of the velocity decreases while
the radial component changes sign. 
In front of the shocks it has a plus sign, meaning gas is accelerated outward,
while behind them it has a minus sign, implying an inward acceleration of the gas.
It must be noted that in the $0.05<r<0.10$ range, the velocity profile is somewhat less clear. This is
due to the inner vacuum boundary condition which reduces the 
number of particles present in this region.

A definitive confirmation of the presence of shocks could be done by checking that the normal 
component of the velocity with respect to the shock shows a transition from supersonic to subsonic
flow. This, however, requires the knowledge of the pitch angle of the spiral shocks, which is
difficult to estimate.
We have thus to limit ourselves to analyse  
the azimuthal and radial Mach numbers.
The mean azimuthal Mach number is around 5 which, if the hydrostatic balance
is valid, would give a disc height to disc radius ratio of 0.2. This is indeed the case as shown below.
On the other hand, the radial Mach number is - except for a few particles -
always smaller than 1. It is however interesting to note that the shock 
clearly reduces the radial component of the velocity, bringing it from 
near supersonic value to zero. 
 
\begin{figure}
  \plotone{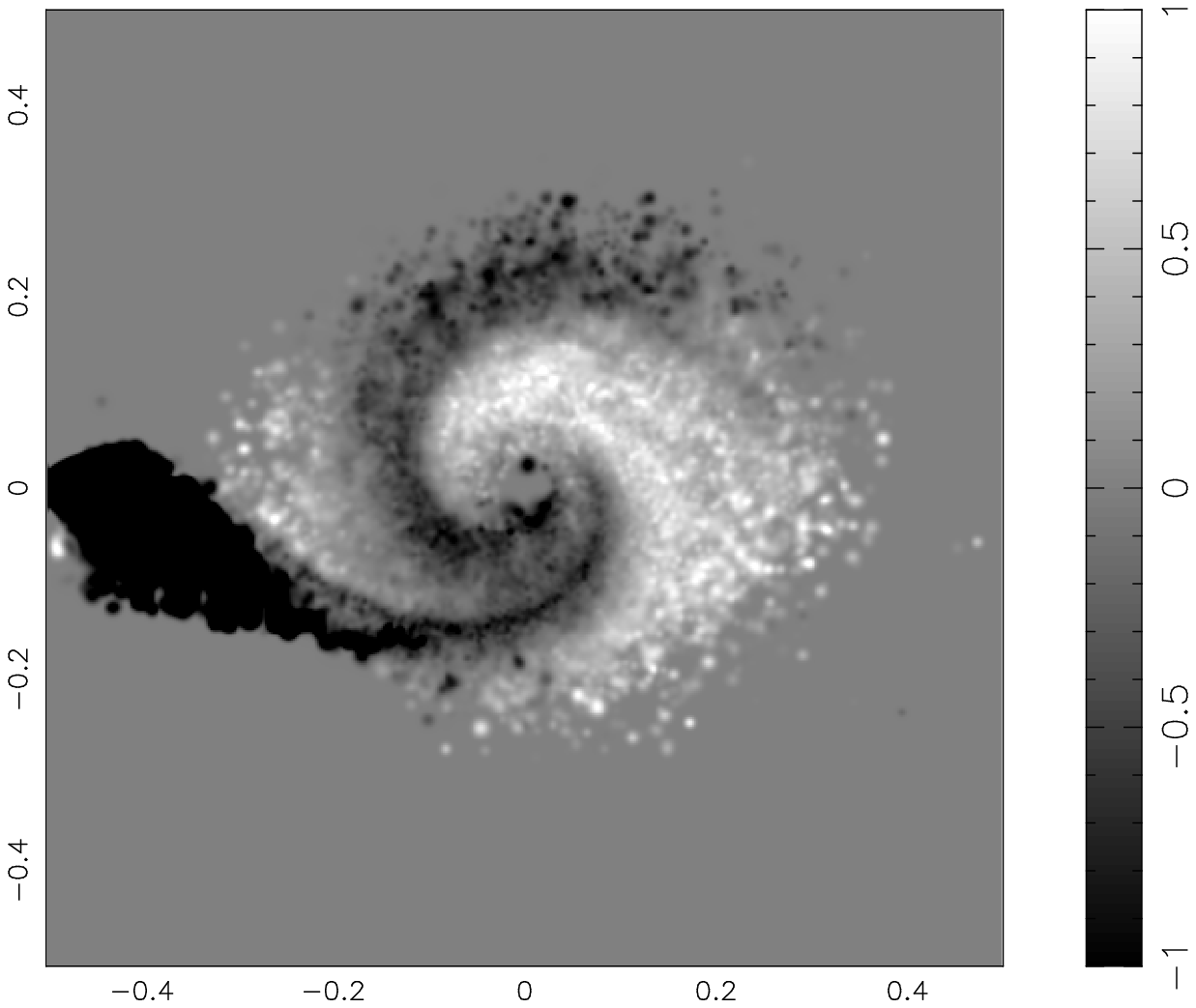}
  \caption{Gray scale plot of the radial Mach number in the  $x-y$ plane
for the $\gamma=1.2$ case at time $T \sim 14$.}
  \label{mac.g12}
\end{figure}

Figure~\ref{mac.g12} is a gray scale plot of the radial Mach number.
The presence of the two-armed spiral clearly demonstrates the effect of the
shock on the velocity. It can also been seen 
that the radial Mach number changes sign along the spiral shock surfaces.

\begin{figure}
  \plotone{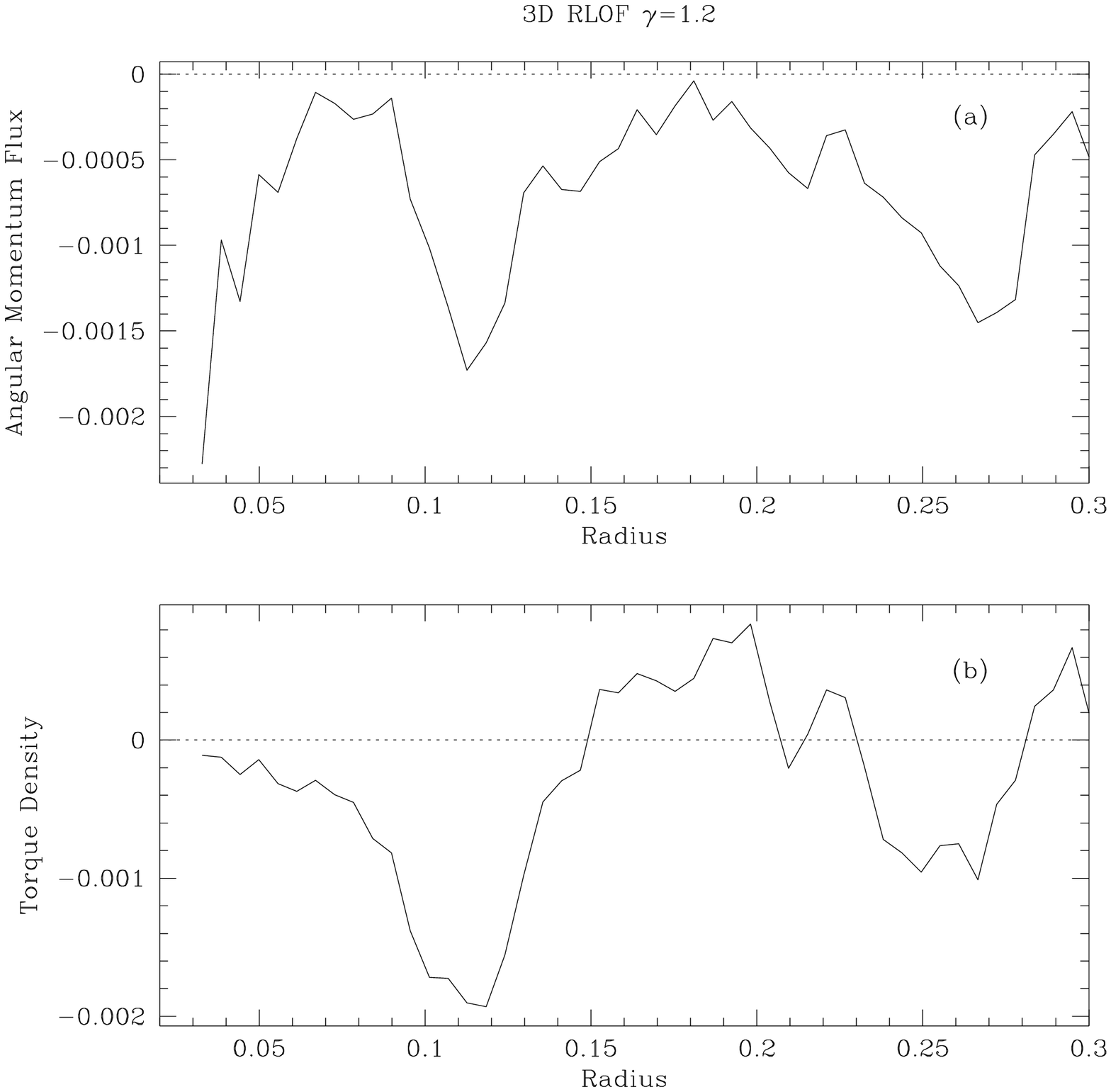}
  \caption{The azimuthal averaged advected angular momentum flux(a) and tidal torque
density(b) as a function of radius for the case $\gamma=1.2$ 
at time $T \sim 14$. The average was done on particles having $|z|<0.02$ only.}
  \label{tr.g12}
\end{figure}

We have also estimated the azimuthally
averaged angular advected momentum flux as well as the torque density between the companion star
and the disc at our final stage for the case $\gamma=1.2$. 
This is shown in Fig.~\ref{tr.g12}.
 
We can see, in the $0.06<r<0.15$ range, 
that the torque is negative, 
meaning that the disc loses its own angular momentum and gives it to the orbital
angular momentum of the binary system. 
For $r<0.06$, the effect of the inner boundary (vacuum) condition is seen, namely that the
gas is accelerated inward.
As the radius of the disc is roughly $0.2$ (see Fig. 3),
the region with $r>0.2$ is not very meaningful to consider. 
However, the same tendency as in the disc region is seen, i.e. the negative torque
density contributes to the angular momentum transfer. 

In  Fig.~\ref{g11xy} and \ref{g101xy}, we show our final results for the $\gamma = 1.1$
and 1.01 cases, respectively.
In both cases, it is more difficult to see any spiral shock although the $\gamma = 1.1$
figure shows a build-up of particles in the disc. These results are rather similar to those
we obtain in 2D (Paper I).
When we consider the mean density as a function of the distance to the primary, the three cases
show a similar pattern : density increases sharply until about $r \simeq 0.1$, then decreases
rather quickly. In the case of $\gamma = 1.01$, at $r=0.2$, the mean density is only 10 \%
of the maximum value, while it is about 40 \% for $\gamma = 1.2$. The maximum value of the
mean density itself is, however, very dependent on $\gamma $ : 1.8, 0.4 and 0.09 for
$\gamma = 1.01$, 1.1 and 1.2, respectively.
 
From Fig.~\ref{g12xz}, we can have an idea of the disc height in the $\gamma = 1.2$ case.
Although a relatively large spread can be seen, the disc appears rather well bound.
We thus believe that even in the case of a large polytropic index, a disc forms.
Using the verified assumption that the particles' distribution in the $z$-direction is a Gaussian,
i.e. $dn(z) ~\sim ~ \exp (-z^2 / 2 H^2) dz$, we estimated the mean height of the disc
as a function of the distance in our 3 simulations. A linear trend was found. The ratio $H/r$
is about 0.09, 0.12 to 0.2 and 0.2 to 0.3 for $\gamma = 1.01$, 1.1 and 1.2, respectively.
Thus, the disc height is about 2 to 3 times larger for $\gamma = 1.2$ than it is for
$\gamma = 1.01$. However, one should note
that in all cases, $H$ almost exactly equals the value expected from the
hydrostatic balance, $c_s / \Omega $. Therefore, our discs appear to be stable.
Of course, inclusion of radiation loss would be required to simulate real discs. This would
reduce the sound speed close to the star and hence make the disc thinner.
Concerning the disc height, we should also mention that we could not observe any variation
with the orbital phase as, for example, Hirose et al. (1991) did. Nor did we observe asymmetry
of the flow for particles outside the orbital plane as reported by Molteni et al. (1991)
and Lanzafame et al. (1992).

Finally, we turn to the accretion rate, shown in Fig.~\ref{acc}. As already stressed
at the beginning of this section, a steady-state has not been reached. This can also
be seen in Fig. ~\ref{acc}. For now, we observe that the mass accretion rate
is slightly dependent of the polytropic index : the smaller $\gamma$ is, the smaller the
accretion rate is. Roughly, the ratio between the injection rate and the
accretion rate is on the order of 50 \% to 60 \%.
Paper I has however shown that the accretion rate is a function of the numerical
resolution. We cannot therefore claim to provide the definitive number.
Also from Paper I, it is not clear whether the mass accretion rate reported here
is entirely due to the presence of the spiral shocks as a large fraction may be
due to the numerical viscosity.
Indeed, in Paper I, we investigated a model in which we did not take into
account the presence of a companion, hence we studied an isolated accretion
disc. In this case, accretion is only due to viscous effect as no tidal torque
is present. However, we noted that the accretion rate so obtained was of the
same order of magnitude as when a companion is present. It is thus difficult
to disentangle the effect due to numerical viscosity and to the companion.
Another problem arises from our inner boundary condition. Here, we simply
remove particles when they enter the boundary. The vacuum so created results in
a spurious force which increases the accretion rate.
We therefore suspect that the difficulty of obtaining a quantitative estimate
of the accretion rate is related to two effects : the SPH articifial viscosity term
and the inner boundary condition. We are presently trying to solve these
problems for further investigations.

\begin{figure}
  \caption{The particle location map in the orbital plane
in the case $\gamma=1.1$ at time $T\sim 19$ : 34779 particles out of a total of 55370 are shown.}
  \label{g11xy}
\end{figure}

\begin{figure}
  \caption{The particle location map in the orbital plane
in the case of $\gamma=1.01$ at time $T\sim 39$ : 23960 particles out of a total of 26375 are shown.}
  \label{g101xy}
\end{figure}

It may, however, be interesting to note that 
the ratio of 50-60 \% is rather similar to the value obtained in our 2D results.
On the other hand, the escape rate is much smaller than the value obtained in the 2D
cases at the same time. 
This can be easily understood : as the gas can now expand in the $z$-direction,
it is more difficult for it to fill the Roche lobe and therefore escape.
 
As we have not yet reached a steady state, this may only be a temporary effect. 
One should also note that, as in the 2D results, the escape rate is a function of $\gamma$.  

\begin{figure}
  \plotone{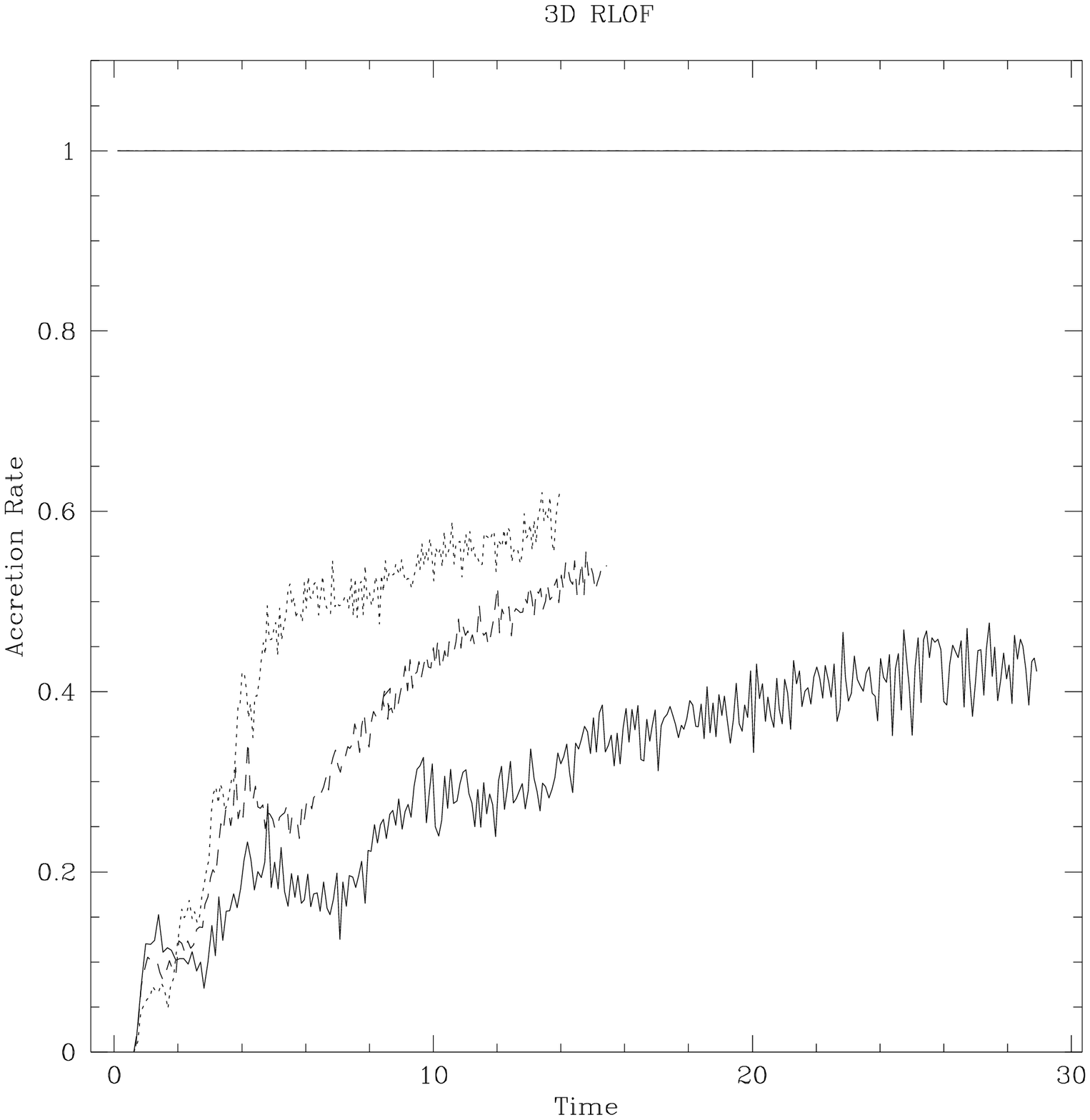}
  \caption{Time history of the  
mass accretion rate for $\gamma =$ 1.01 (full curve), 1.1 (broken curve) and 1.2
(dotted curve). The unit is the mass injection rate, 0.00016.}
  \label{acc}
\end{figure}
 
\subsection{Keplerian disc}
\begin{figure}
  \plotone{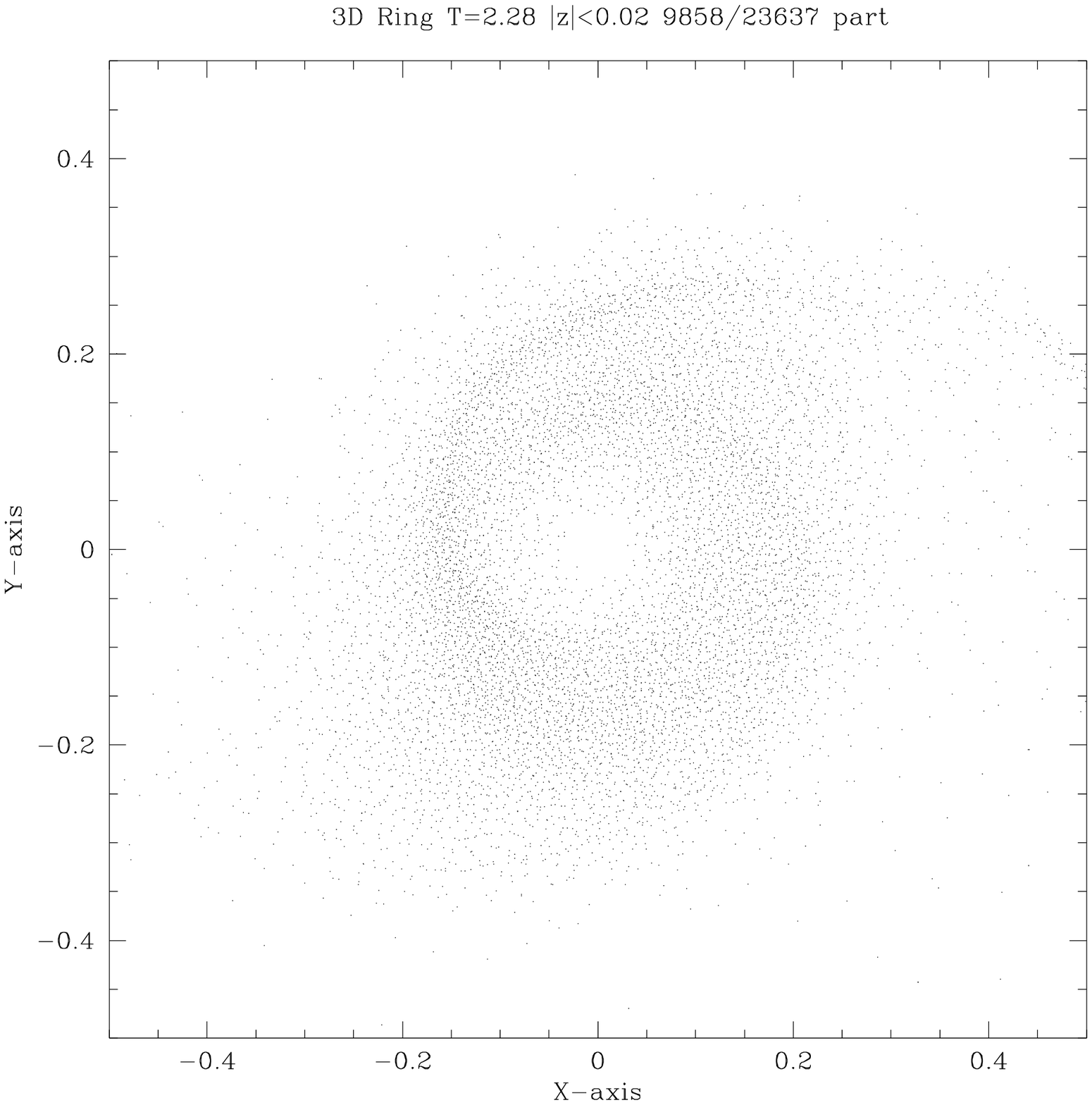}
  \caption{The particle location map on the orbital plane
in the case of the Keplerian disc simulation with $\gamma=1.2$ at time $T\sim 2.3$}
  \label{kepl12xy}
\end{figure}

Figure \ref{kepl12xy} shows the particle distribution in the orbital plane in the case
of no mass inflow.
Although the simulation could be followed for only a short time, 
the spiral shock is again clearly seen, as well as the elongation of the disc
in the $y$-direction. 
From a close inspection of the particle distribution, 
it appears that the spiral pattern does not change in the $z$-direction. 
We also estimated the disc height using the verified assumption of a Gaussian distribution.
The mean disc height, $H$, is phase independent and scales almost linearly with the 
distance to the primary star in the equatorial plane, $r$, in the following way :

\begin{equation}
H = 0.13 ~ r + 0.011.
\end{equation}

Although this shows that $H$ may take values between about 20 and 35 \% of $r$, it 
almost exactly equals the value expected from the hydrostatic balance. Thus, in this case
also, the disc height is only due to the pressure. 
Pressure is also responsible for the spread in velocity around the Keplerian value.
We can see no difference in the disc height or in the velocity distributions of this simulation
and the one described in the previous section. We can therefore not confirm 
the claim of Hirose et al. (1991) that the interaction between the L$_1$ flow and the disc is responsible
for a large and spatially-variable disc height. We recall that Hirose et al. (1991) did not
include pressure in their calculations.

\section{Conclusions}
We have presented our results of 3D SPH calculations of accretion discs in 
two different situations, Roche lobe overflow and the evolution
of a Keplerian disc.
The results of our simulations clearly show the following:

(1) Even in the $\gamma=1.2$ and $\gamma=1.1$ cases, discs do exist.
Their height, which depends only on the radial distance, is rather 
large but totally consistent with the hydrostatic balance equation. 
Compared with other authors, 
we use the SPH method (which includes the effect of pressure), with 
about ($5 - 6)\times 10^4$ particles and a variable smoothing length.

(2) In the $\gamma=1.2$ case, spiral shocks are seen.
These spiral shocks have a clear effect on the velocity of the gas.
In the $\gamma=1.1$ and $\gamma=1.01$ cases, although spiral shocks are
not clearly seen, a build-up of density is present. These results are
in agreement with our 2D results presented in Paper I.
This gives strong support to the accuracy of 2D simulations.

(3) Spiral shocks were also observed in discs without 
mass supply from the inner Lagrange point. In this case, the structure
of the disc is rather similar to that obtained in the Roche lobe overflow 
simulation. The impact of the incoming stream seems therefore to have
a rather small effect on the structure of the disc.
One may note that in our simulations the width of the stream at the 
inner Lagrange point is always smaller than the height of the disc.
This, combined with the fact that we use a mass ratio of unity, may 
explain this conclusion.

(4) Concerning mass accretion rates, although we could not reach a definite conclusion
due to the long time required to attain a steady state, we observe that the
ratio between the mass accretion and mass injection rates is about 0.5 to 0.6.
This is however very dependent on the numerical viscosity  and the inner
boundary condition.

\section*{Acknowledgments}

It is a pleasure to thank an anonymous referee for suggestions to an 
earlier version of this paper.
The calculations were mainly performed on Fujitsu VX/4R 
at the National Astronomical Observatory, Japan. 
TM is supported by the Grant-in-Aid for Scientific Research (05640350,
07640410, 08640375) of the Ministry of Education, Science and Culture
in Japan.
HB beneficied from an EC-JSPS fellowship.
Part of these simulations have been done using computers in the Yukawa Institute
(Kyoto University) and in the National Astronomical Observatory of Japan (Tokyo).

\bsp

\end{document}